# Blockchain-Based Trusted Achievement Record System Design

Bakri Awaji
School of Computing
Newcastle University
Newcastle, UK
b.h.m.awaji2@ncl.ac.uk

Ellis Solaiman
School of Computing
Newcastle University
Newcastle, UK
ellis.solaiman@ncl.ac.uk

Lindsay Marshall
School of Computing
Newcastle University
Newcastle, UK
lindsay.marshall@ncl.ac.uk

## ABSTRACT
The primary purpose of this paper is to provide a design of a blockchain-based system, which produces a verifiable record of achievements. Such a system has a wide range of potential benefits for students, employers and higher education institutions. A verifiable record of achievements enables students to present academic accomplishments to employers, within a trusted framework. Furthermore, the availability of such a record system would enable students to review their learning throughout their career, giving them a platform on which to plan for their future accomplishments, both individually and with support from other parties (for example, academic advisors, supervisors, or potential employers). The proposed system will help students in universities to increase their extra-curricular activities and improve non-academic skills. Moreover, the system will facilitate communication between industry, students, and universities for employment purposes and simplify the search for the most appropriate potential employees for the job.

## 1. INTRODUCTION
### 1.1 Motivation
Every individual studying in higher education (tertiary education leading to the award of an academic degree [21]); must have a record of their learning and progress at university. The higher education system takes a unified approach in the creation of these records, producing official transcripts which authenticate students' academic achievements. An official university transcript provides an individual with proof of their performance [21]. Official transcripts are crucial records for an individual's employment, enabling potential employers to assess a candidate's relevant education. Employers might also assess a candidate's ability by requesting a submission of a work portfolio. Studies have shown that adequate records of service-based or project-based learning significantly increase employment opportunities [8][20]. However, there is no guarantee that work presented in a portfolio, or that transcripts or certificates are the applicant's own unless these can be recorded with the aid of a trusted achievement recording framework.

There are many online resources which enable users to construct records of achievements through various mediums – from providing different structures and styles to the utilization of social networking platforms such as LinkedIn or Facebook. However, it is well known that many job-seekers exaggerate their achievements, and currently there are no adequate methods in place, enabling employers and recruiters to validate an individual's claimed achievements [4]. A survey conducted by Higher Education Degree Data (HEDD) [8][9] found that approximately 30% of students and graduates embellished or exaggerated their academic achievements. This is an increasingly prevalent problem employers currently face. NGA HR services, which is a provider of human resources business process outsourcing (BPO) services and consulting in the United Kingdom released a statement in 2018[7], revealed that "90% of HR directors have found exaggerations on a job application."

Additionally, research [5][7] suggests that "the average organization is spending up to £40,000 a year on confronting these challenges". Fraudulent records of achievement pose a problem not only for employers but for honest, qualified candidates who cannot compete with dishonest candidates. The development of measures for the prevention of certification fraud is vital. A study conducted in 2018 by the NGA identified some of the most common areas in which candidates were dishonest about their achievements. The study suggests that 44% of candidates were dishonest in reporting their skills, while 43% were dishonest about their career history. Furthermore, 39% of candidates were found to be dishonest about professional qualifications and 32% to be dishonest about education qualifications. 27% of candidates lied about membership to an industry body, and 24% provided false references [10].

### 1.2 Benefits of the proposed system
A blockchain-based achievement record system which produces a verifiable record of achievements has a wide range of potential benefits for students, employers and higher education institutions. This system would enable students to present their academic achievements to their employers enhancing their employability. An official transcript also offers students the ability to review their progress and plan for their future accomplishments, both individually and with support from other parties, helping them to increase their extra-curricular activities and improve non-academic skills. Furthermore, the knowledge that all of their academic achievements will be recorded on a transcript may motivate a student to maintain a strong record of work and to perform well in assessments, adding value to their experience of higher education. In terms of benefits to the education system itself, a trusted system for producing records of achievements helps to reduce administrative tasks and is efficient and cost-effective. It also has the potential of contributing to enhancing student admissions quality standards, providing transparency as to the student's achievements outside and within their formal curriculum. The benefits of such a system for an employer include the production of trustworthy and verified achievement records which have a standardized template to ensure validity. A transparent and detailed picture of a candidate's achievements within higher education is helpful for employers when selecting and hiring graduates [1].

## 1.3 Aim

This project will investigate a method for providing stakeholders (for example students, employers and educational organizations) with a system for the organization and validation of certificates and other relevant data in order to construct a trusted record accessible to multiple parties (for example students, university admissions, student registration services, and potential employers). This form of trusted record would permit admission staff to evaluate an applicant and to determine whether they meet university requirements without the necessity of contacting the certificate issuer for validation. For example, this kind of trusted record would be useful for medical students and medical employers, providing evidence of required training. Additionally, such trusted records would enable employers to verify information given on a CV while providing a graduate with a single comprehensive record of achievements. Institutions would also be able to create a verified and comprehensive picture of their students' achievements; not only their academic work but their extra-curricular activities, prizes, employability awards and voluntary experience and positions a student has held in student clubs or societies. One of the principal motivations for this project is recent research and development of innovative distributed ledger (blockchain and Smart Contract) technology, which provides a platform on which trusted; authentic records can be generated. Such technology would ensure that CVs and data records are authentic, queryable, tamper-resistant and that they have not been subjected to any form of forgery.

## 1.4 Objectives

The principal aim of this system is to use blockchain and smart contract technology to build an auto-create achievement record system for students to:

- Record all the activities that student performs, academic or non-academic during their life at the university.
- Provide proof of students' qualifications and skills that are not recorded in their official academic transcripts and certificate.
- Connect students to industry via a trusted network that facilitates the hiring process for employers. Hence, they can search for potential employees among students in the system; whose skills and qualifications match the job/position requirements.
- Encourage students to learn and improve their skills and earn knowledge from various resources besides academic courses and modules.
- Help students to build a future learning plan that keeps up with their future career aspirations.

## 2. BACKGROUND
## 2.1 Blockchain

Blockchain [5] technology facilitates a continuously growing list of records, called blocks, where each block contains a hash of the previous block. Each block is linked and secured using cryptography. Each block in the blockchain contains data, as well as an index, a timestamp, a hash of itself, a hash of the previous block. Therefore, it is necessary to define the various parts central to blockchain technology.

- Block: The fundamental element within a blockchain is the block. Each block contains data and a hash of the previous block. It is this hashed pointer that ensures the integrity of the data and makes a list immutable. The collection of data is generally capped at an upper limit, which is determined by either the size or the number of data units.

- Chain: Blocks form a chain because they are linked together by an algorithm which utilizes the information contained within the previous block.

- Node: A system utilized within the blockchain for the verification of the authenticity of each block. Nodes also maintain the digital ledger, which ensures the blocks are kept in chronological order.

## 2.2 Smart Contract

The term was first coined by Szabo in the 1990s [7]. A smart contract is an event-condition-action stateful computer program. It can be created on-top of blockchain technology in order to implement a distributed application for multiple parties who do not fully trust one another [10][21]. It can be defined as self-executing code that implements the agreed rules and conditions of interaction between the parties. Smart contracts aim to automate and replace the role of a TTP (trusted third party), which normally mediates interactions between the players within a distributed application. Therefore, benefits include the reduction of third-party costs. Also, with the use of blockchain technology, interactions are much more secure. A more detailed overview of key concepts related to smart contracts and their implementation can be found here [18][13][14]. The state of a contract is comprised of the storage and balance of a contract. This state is stored on the blockchain, subject to updates every time the contract is put into effect. Each smart contract is assigned with a unique address, which in the Ethereum blockchain for example has a size of 20 bytes [2]. Once a contract has been deployed in the blockchain its contract code is unalterable. In order to execute a smart contract, a user sends a transaction to the contract's address. The transaction that is sent by the user will be executed by each consensus node (miner) within the network until consensus (agreement) on the transaction's output is achieved. Once consensus has been achieved, and the transaction is executed, the state of the application is updated appropriately within the blockchain. A smart contract is capable of many functions. For example, it can read/write to its private storage, store money within its account balance, send and receive money and messages from other contract users or other contracts. Additionally, the contract is even capable of creating new contracts. Smart contracts can be either deterministic or non-deterministic [17]. While a deterministic contract does not require information from parties external to the blockchain in order to run, a non-deterministic contract differs in that it depends on information (called oracles or data feeds) which derives from external sources. An example of such a non-deterministic contract would be a contract which requires information on the current weather in order to run information which is not included within the blockchain [21].

## 3. BLOCKCHAIN APPLICATIONS WITHIN EDUCATION

Blockchain applications currently, are mainly in the experimental stage within higher education [20]. Examples include their introduction in order to provide degree management support and to aid in summative evaluations[17]. Blockchain can also aid formative evaluation processes within higher education. It is useful in tracking students' educational development as well as the methods by which learning is implemented. The University of Nicosia was the first higher education institution to utilize Blockchain technology in order to manage students' certificates received from MOOC platforms [17]. The Massachusetts Institute of Technology (MIT) has since developed a credentials system that enables greater control over students' certification without reliance on third-party intermediaries (universities or employers) in order for the effective storage, verification and validation of credentials to take place called Blockcert [5][16]. The Blockcerts open platform [5] for digital certificates and reputation has been developed through combining Blockchain technology with reliable cryptography. Another institution which has proposed the use of Blockchain technology is the Knowledge Media Institute (KMI) in the Open University. KMI has proposed the Open Blockchain project. The objective of the Open Blockchain project is the creation of a network which acts as a spearhead for all Blockchain projects within higher education. KMI intends to enhance the standards around badging, certification and reputation online, utilizing the blockchain as a trusted ledger [11]. Additionally, Sony Global Education has used Blockchain technology in developing a global assessment platform for the storage and management of degree evidence [12]. Finally, Turkanovic [19] has put forward a proposal for a decentralized global platform with a basis in Blockchain technology, which would be used for the management, storage, and sharing of digital certificates called the EduCTX platform. Experimentation with and utilization of Blockchain technology within higher education has been underway for some time [19]. A systematic mapping study of blockchain applications in higher education can be found in [3].

## 4. MOTIVATION SCENARIO

### 4.1 Initial Scenario

The following initial scenario involves four different actors. The first actor is the system administrator, or sysadmin, responsible for the execution of the smart contract on the blockchain and the registration of universities on the system. The second actor is the university or learning institution, responsible for the authentication of student records. The third actor is the student, who utilizes the system in order to create a record of their achievements. The fourth actor is an employer, who utilizes the system in order to validate the candidate's certification and to assess candidates with the use of their records of achievements.

Each of these actors, or users, interact with the system in different ways, according to a sequence. This is described below:

1. The sysadmin executes the smart contract on the blockchain. Following deployment of the smart contract on the blockchain, the sysadmin can register universities into the smart contract's storage. This function is one which can only be executed by the sysadmin of the smart contract, or, in other words, executed from the address from which it was deployed.
2. The university name and address are stored in the smart contract, utilized for the authentication of certificates.
3. The sysadmin registers the university in the frontend of the system (meaning that the university name is added to the list of the universities registered in the smart contract).
4. The university adds its students to the system.
5. Students added to the system now have access to the system with the use of an ID number.
6. Each time a university seeks to authenticate certificates, the

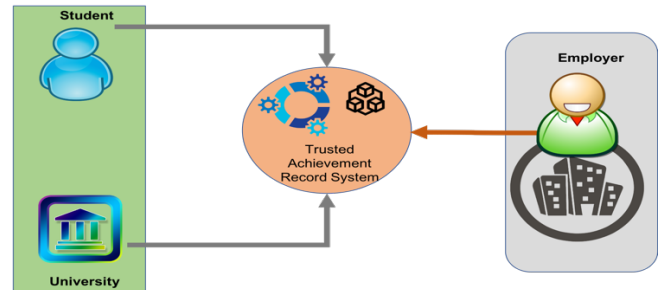

**Figure 1. Motivating Scenario.**

relevant student is selected from the list.

7. Following this, the relevant documents are uploaded to the frontend, which hashes the certificate and calls to a function on the smart contract to ensure this hash is stored on the blockchain.
8. Following authentication of a student certificate, the certificate is automatically added to the student's record of achievements on the system.
9. When this occurs, the UI also sends an email to the student with a hash of the certificate.
10. Once given this hash, the student can send this certificate to their employer(s).
11. Once the employer receives this hash, they upload it to the UI for verification.
12. The frontend calls on a function in the smart contract in order to verify that the hash was authenticated by the university.
13. The true/false boolean result is then displayed on the frontend, in the verifying window.

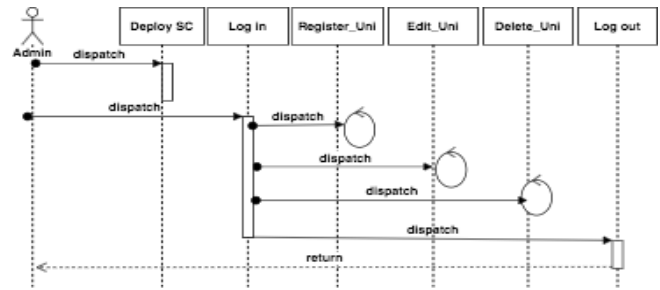

**Figure 2. Service Level sequence diagram/ Admin.**

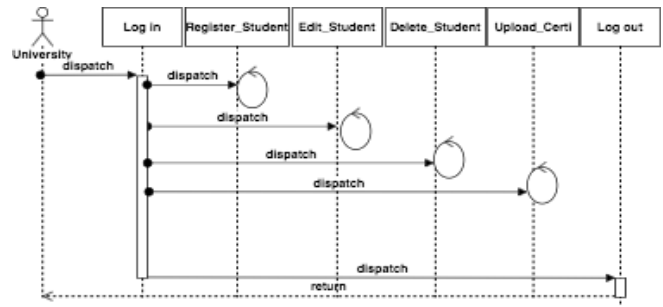

**Figure 3. Service Level sequence diagram/ University.**

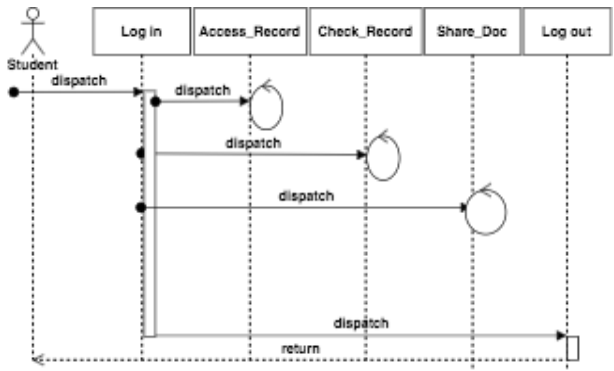

**Figure 4. Service Level sequence diagram/ Student.**

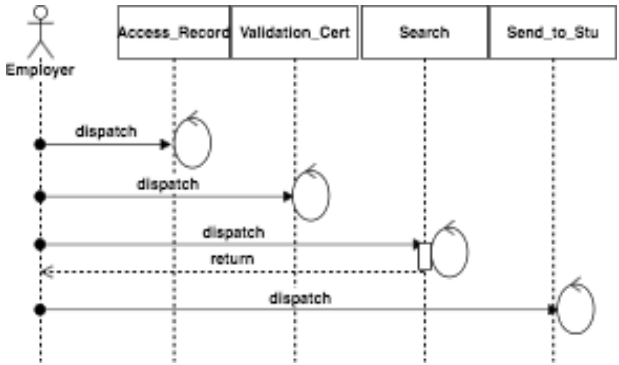

**Figure 5. Service Level sequence diagram/ Employer.**

This motivating scenario demonstrates how data flow occurs within two different system platforms; it occurs off-chain in the frontend application and on-chain on the blockchain in the backend of the system.

## 4.2 Scenarios for Analyzing Special Requirements

The system must be designed in a way which makes it capable of coping with unexpected issues caused by the system itself or its users. Therefore, one primary necessity is the creation of scenarios for potential issues in order to pre-empt issues and to formulate plans for their resolution. These issues might include faults within Blockchain behavior, user login, and permissions.

### 4.2.1 Faults related to blockchain.

One potential fault lies in the use of Ethereum Blockchain, which relies on miners for the verification of transactions, management of the network, the issuing of new Ethereum tokens and for securing the network. With an Ethereum transaction, the address of the smart contract is encrypted on the blockchain, alongside the certificate hash, the gas and the private key authorization for verification of the side transaction. Ethereum miners will pick up transactions with the highest fees, meaning that miners will more quickly verify transactions with a high amount of gas. Likewise, it will take longer to achieve confirmation for a transaction with a low amount of gas. Additionally, enough Ether must be contained within the senders' digital wallets in order for the mining fee to be paid to the miners. This indicates that a scenario must be created in order to plan for the event of a "transaction not confirmed (verified) on the blockchain". However, this research does not intend to examine the technical faults relevant to blockchain or the consensus process.

### 4.2.2 Identity Authentication Issues.

Any potential issues surrounding user authentication must be explored, as the privacy of the data is crucial. Only authenticated users should have access to information, and not all users should have access to sensitive information. This means that a piece of information must be given in order for a user to confirm their identity. Despite this measure, some issues with identity authentication are expected. In order to counter these issues, the system was designed to give users the ability to reset the information they use to access the system via a trusted and secure process.

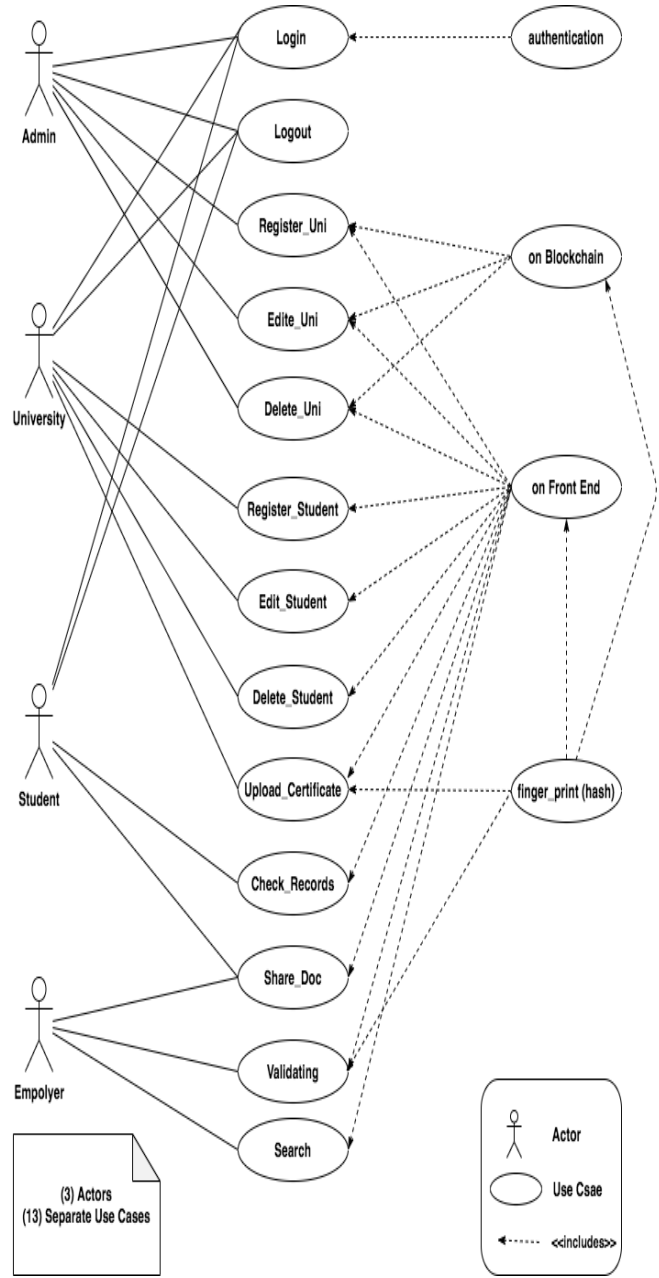

**Figure 6. System Use Case Model.**

## 5. SYSTEM STRUCTURE

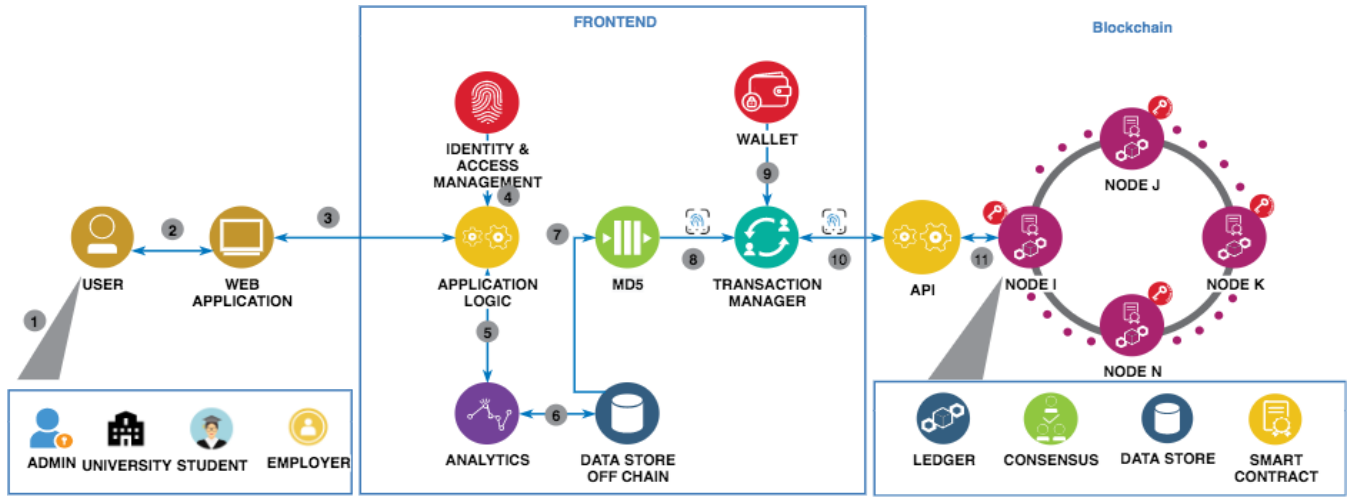

**Figure 7. The System Structure.**

Figure 4 illustrates the overall system design, and covers the previously discussed requirements and components including a frontend and backend. This architectural design means that there are two ends with separate set dependencies (libraries and frameworks). While the frontend acts as a presentation layer, being what the end-user sees upon entering the site, the backend provides the data and logic which enables the frontend to function.

### 5.1 Blockchain

This system is a Blockchain-based application. Blockchain has been selected and is considered an adequate technology as it fulfils the specifications required for the verification of educational certificates. Several facts have influenced this reasoning:

- The need for manual verification of transactions is removed, as all information is verified by a decentralized network of computers.
- Information is permanently stored in the blockchain, removing the risk of deletion and thus the necessity for additional security services.
- Falsification or modification of transactions on the blockchain cannot take place. The hash system is used in order to verify certificates, and no user is capable of modifying this information or of uploading a false hash into the network.

The motivating scenario included in Section 3 justifies the storage of data on the blockchain. An MD5 hashing algorithm will be utilized to encrypt the data at the frontend of the system, increasing levels of security and ensuring the privacy of system end-users data. Furthermore, the MD5 hashing algorithm evades the complexity of consensus mechanisms and is capable of validating blocks on the private and consortium Blockchain types. Thus, a public Blockchain has been selected as it is the most appropriate type to utilize within this system.

The following is an explanation of the sequence of events which take place when data is added to a Blockchain:

1. A request for the exchange of data is given by a sender and received by one of the nodes in the blockchain.
2. The receiving node broadcasts the incoming data to other nodes, adding it to the current transaction pool.
3. When the block's limit is reached (determined by either the size or number of units), the nodes begin mining the block.
4. The nodes compete to identify the found proof of work solution. When one of the nodes succeeds in mining, the solution is broadcasted to the other nodes.
5. The other nodes are responsible for verifying the output and assessing its validity before all blocks in the chain are verified, and the newly mined block is added.

### 5.2 Smart Contract

The smart contract plays a role which is of primary importance within the system – connecting the blockchain with the frontend. Within this platform, the smart contract removes the necessity of human management and thus significantly reduces the risks of documentation fraud for educational institutions and employers. The smart contract executes actions automatically on the frontend with use of API connectivity.

The primary programming language for the writing of smart contracts within Ethereum Blockchains is solidity. Solidity is a contract-oriented language, meaning that the responsibility for secure storage of programming logic in a transaction with the blockchain is the smart contract's responsibility. Solidity is not only a contract-oriented language, but it is also high-level. It is used in the writing, design and implementation of smart contracts and it is designed so as to run on the Ethereum Virtual Machine (EVM). The EVM is hosted on Ethereum Nodes which are connected to the blockchain.

### 5.3 Hash Algorithm

The MD5 hash algorithm [6] has been selected to be used on this platform due to MD5's (message-digest algorithm) ability to produce "fingerprints" using one-way functions. MD5 hash algorithms reduce data's heterogeneity, mapping something which has many parts down to just a few parts (128 in the case of MD5), and thus making collisions as rare as possible. Additionally, MD5 was used because it enabled us to compare and store small hashes, which is more accessible than with entire original sequences. One-way hashes are utilized in cryptography in order to enable verification without revealing the original information. MD5 hashing was used in this system in order to verify data's authenticity without necessitating data encryption beforehand. This hash is vital to this project because the documents can only be stored in the blockchain once they have been converted into hashes; only hashes will be stored on the blockchain.

## 6. CONCLUSION

In conclusion, this research has aimed to demonstrate that the use of Blockchain technology for the certification and verification of achievements in higher education has great potential in the global market. It is a method which would be sustainable and advantageous for multiple parties. Credential fraud is widespread and pervasive, having a negative impact on educational institutions, impairing social development, and effecting significant economic losses. Current solutions, such as legacy credential verification systems, are clumsy and are not time nor cost efficient. In addition to this, they lack efficacy in their response to corrupt practices, such as fraud on the part of educational institutions and accreditation bodies. The record of achievements that is proposed with the use of Blockchain technology is comprehensive in tackling widespread fraud. This system is significantly improved in comparison to legacy systems, being both more user-friendly and more efficient. The Blockchain technology method is a solution which effectively integrates into the existent credential verification ecosystem. Furthermore, it offers a novel on-chain credential revocation mechanism which does not require either students or employers to maintain cryptographic credentials. This work aspires to positively contribute to the ongoing efforts towards the prevention of credential fraud.